\documentclass[12pt]{iopart}
\usepackage[final,hiresbb]{graphicx}
\usepackage{indentfirst}
\usepackage[usenames]{color}

\begin{document}

\title[Embedded Ribbons of Graphene Allotropes]{Embedded Ribbons of Graphene Allotropes: An Extended Defect Perspective}
\author{David J. Appelhans, Lincoln D. Carr and Mark T. Lusk}
\address{Department of Physics, Colorado School of Mines, Golden, CO 80401, USA}
\ead{mlusk@mines.edu}

\begin{abstract}
Four fundamental dimer manipulations can be used to produce a variety of localized and extended defect structures in graphene. Two-dimensional templates result in graphene allotropes, here viewed as extended defects, which can exhibit either metallic or semiconducting electrical character. \emph{Embedded allotropic ribbons}--i.e. thin swaths of the new allotropes--can also be created within graphene. We examine these ribbons and find that they maintain the electrical character of their parent allotrope even when only a few atoms in width. Such extended defects may facilitate the construction of monolithic electronic circuitry.
\end{abstract}

\maketitle

\section{Introduction}
\label{sec:introduction}

The presence of defects in condensed matter has a dramatic influence on electrical character. Extended defects such as dislocations, grain boundaries, and phase interfaces, for instance, tend to have a deleterious effect on electron mobility and can serve as unwanted recombination sites in photovoltaic materials. The introduction of localized dopant defects to tune electronic structure, on the other hand, is a mainstay of the semiconductor industry.  Whether unavoidably present or intentionally introduced, localized and extended defects tend to be statistically distributed. The precise placement of defects, though, crosses the line into materials architecture.
\begin{figure}[hptb]
\begin{center}
\includegraphics[width=.7\textwidth]{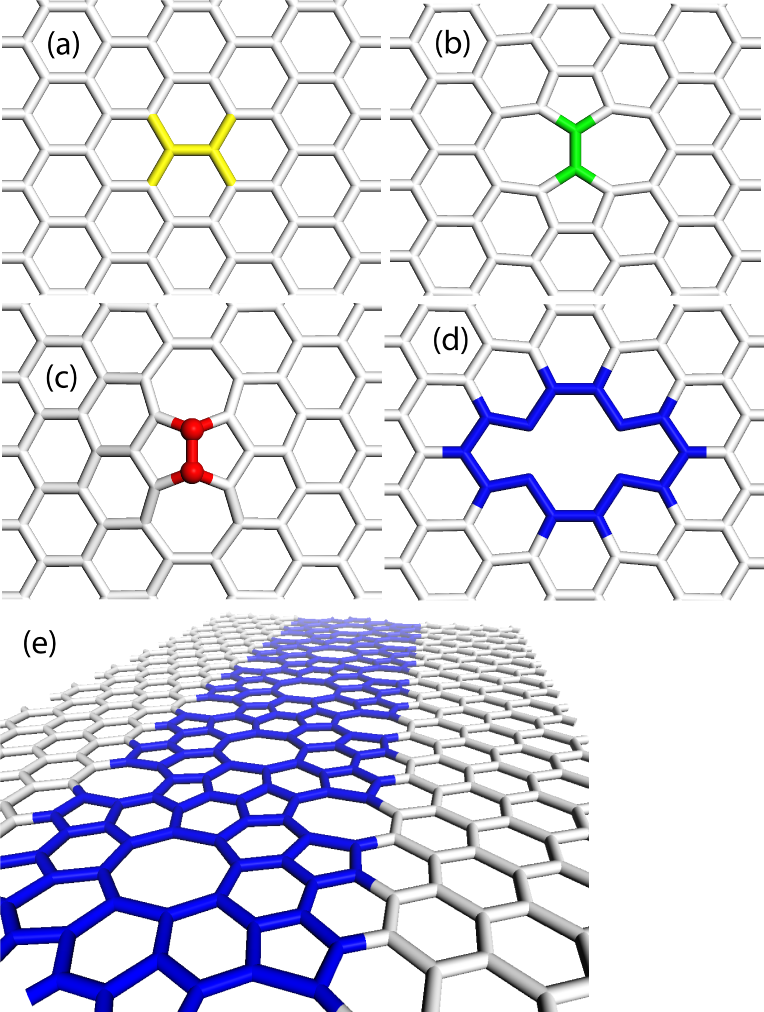}
\caption{
A four-letter alphabet for defect creation: (a) identity operation leaving dimer as is; (b) dimer rotation to create STW defect; (c) dimer addition to create ISTW defect; (d) dimer removal to create DV defect; (e) extended defect, a semiconducting ribbon created with a combination of dimer rotations and dimer removals.}
\label{Alphabet}
\end{center}
\end{figure}

Graphene offers a particularly attractive setting for such \emph{defect engineering}. Its two-dimensional geometry implies that the entire domain is accessible to external manipulation.  Although such manipulations may seek to modify thermal, mechanical, or chemical character, our current focus is to provide a means for the creation of monolithic carbon logic circuitry. This requires that it be possible to create conductors, semiconductors, and charge confinement structures. In addition, composite systems must be structurally stable and should not unduly modify individual character.  Charge may then be introduced, confined, gated and routed within a single sheet.

Our approach is in contrast to a number of other strategies which are being explored to manipulate the electronic structure of graphene. Those tend to focus on the use of external agencies to distort graphene so that it exhibits a band gap. Such external agencies include the use of strongly interacting substrates~\cite{Ohta2006, Giovannetti2007, Ciobanu2009} or the application of electric fields in graphene bi-layers~\cite{Ohta2006,  Zhang2005,  Nilsson2007}. Quantum confinement effects can also be exploited as in carbon nano-ribbons (CNRs)~\cite{ Brey2006, Mucciolo2009} or carbon nano-meshes~\cite{Bai2010}. Another approach to tailoring graphene is chemical modification. For instance, hydrogen can be introduced so as to change the bonding from sp-2 to sp-3 to create the corrugated insulator graphane~\cite{ Sofo2007}. Similarly, a graphene oxide insulator can be created via bonding with hydroxyl groups~\cite{ Dikin2007}. The external agency can also appear in the form of neighboring regions of foreign material as in composite two-dimensional structures that harbor domains of graphene~\cite{ Ci2010}. We consider an intrinsic approach to modification, though, wherein defects are introduced into the graphene in order to endow it with new electronic properties. Several other researchers in the field are also exploring such defect modifications~\cite{Rutter2007}.

In a series of earlier works, we have identified simple defects from which a variety of localized and extended structures can be fabricated~\cite{LuskCarr2008, LuskCarr2009, LuskCarr2010, CarrLusk2010}. Our perspective is now that the design space is best described as four basic operations that can be performed on a single carbon dimer. These are shown in Fig.~\ref{Alphabet}. The first is the identity operation in which an existing carbon dimer is left unchanged. The second is the familiar Stone-Thrower-Wales (STW) defect consisting of a rotation of a carbon dimer. Addition of a dimer constitutes the third configuration and results in an Inverse Stone-Thrower-Wales (ISTW) defect~\cite{Sternberg2006, LuskCarr2008}. Conversely, removal of a carbon dimer results in a di-vacancy (DV) defect, the fourth letter in our alphabet. All structures to be considered can be reduced to a weighted sum of these basic building block. Even when distinct from the actual synthesis routes, the analytical fabrication of structures as a sequence of dimer additions, subtractions, rotations and preservations lends a new type of insight into form and function.

All elements of this defect alphabet have been experimentally observed~\cite{Banhart2009, Hashimoto2004, Urita2005, Batzill2010}. STW defects have been extensively studied in both graphene~\cite{Li2005, Stone1986} and CNTs~\cite{Kotakoski2006, Kim2006}, and they can be created by scanning tunneling microscopy (STM)~\cite{Berthe2007} or atomic force microscopy (AFM)~\cite{Sugimoto2005}. DV defects can be precisely created via irradiation~\cite{Hashimoto2004}. Recently, an electron irradiation beam focused to $1~\AA$ was shown to be capable of precisely creating vacancies in CNTs~\cite{Banhart2009}. DVs have also been considered theoretically within a CNT setting, where molecular dynamics annealing causes them to reconstruct into 5-8-5 defects~\cite{Yuan2009, Sammalkorpi2004}. A series of \emph{ad-dimer} ISTW defects have been experimentally achieved~\cite{Batzill2010} by focusing on the interatomic spacing of the defect rather than on physically depositing a dimer on the graphene. Direct dimer addition may be another avenue of achieving the ISTW defect as advances in experimental methods of graphene manipulation are now approaching atomic fidelity using Scanning Tunneling Microscopy (STM) \cite{Berthe2007}, Atomic Force Microscopy (AFM) \cite{Sugimoto2005} and via novel inventions such as the one proposed by Allis and Drexler \cite{Allis2005}.

\begin{figure}[ptb]\begin{center}
\includegraphics[width=.8\textwidth]{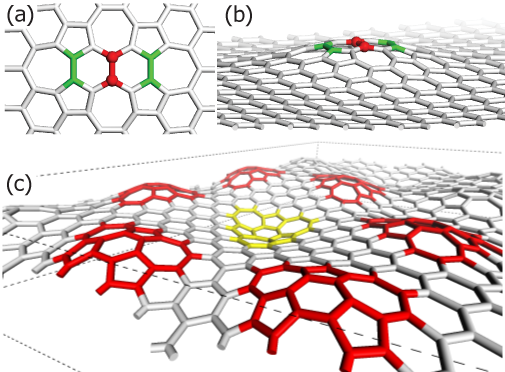}
\caption{
A metacrystal created by patterning a localized defect structure: (a) an ISTW defect with STW defects on either shoulder relaxes into a non-planar blister; (b) perspective view of the blister shown in (a); (c) patterning of blisters results in a metacrystal. Blisters with both positive polarity (red) and negative polarity (yellow) are possible. (Figure (c) reproduced from~\cite{LuskCarr2008}.)
}
\label{metacrystal}
\end{center}\end{figure}

In this work, we focus on patterned defect structures obtained by replication of a defect unit, a combination of the basic alphabet members. For instance, ad-dimers can be used to create nonplanar blisters by adding STW defects to either side as shown in Fig.~\ref{metacrystal}(a, b). Larger blisters can also be created using additional ISTW defects, and graphene islands can even be made to bubble up when corralled by a loop of the ad-dimers~\cite{LuskCarr2008}. Replication of localized blisters results in the metacrystal shown in  Fig.~\ref{metacrystal}(c). However, even closer packing can result in planar allotropes of graphene~\cite{LuskCarr2009}.  These allotropes may be either conducting or semiconducting merit a brief review because they are a starting point for the embedded defect ribbons that we have in mind,  i.e.,  thin strips of an allotrope existing within a parent graphene sheet.  These ribbons are the monolithic analog of graphene nanoribbons~\cite{Brey2006}, but their electronic character is intrinsic and not due to the presence of free boundaries. We will show that both conducting and semiconducting ribbons can be constructed within a completely planar setting.

\begin{figure}[hptb]
\begin{center}
\includegraphics[width=0.7\textwidth]{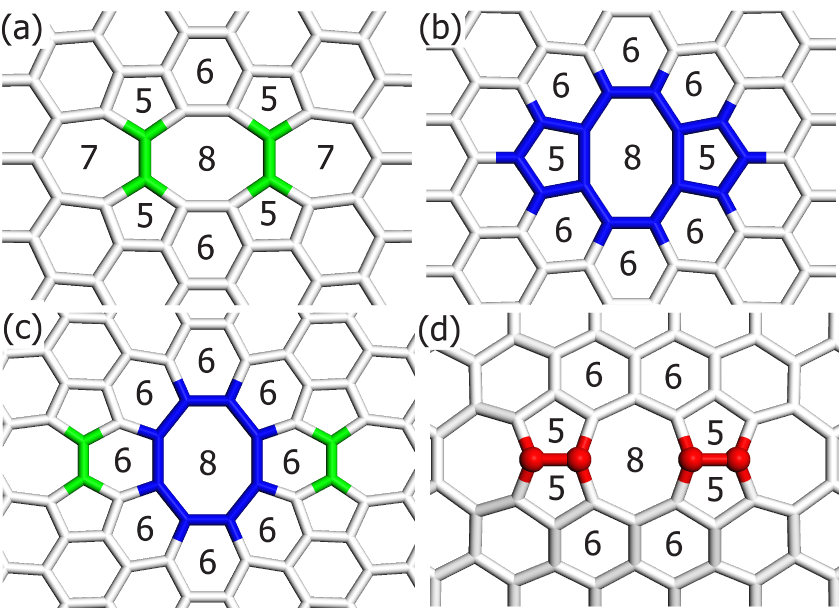}
\caption{
Octagonal structures can be created using any of the fundamental defect building blocks: (a) DV;  (b) DV with two STW defects; (c) two STW defects; (d) two ISTW defects. Structures shown in (a), (b) and (c) are planar while that shown in (d) is nonplanar.}
\label{Octagon}
\end{center}
\end{figure}

The electrical, and most likely magnetic, properties of graphene allotropes are not attributable to any one of the fundamental dimer operations but rather the type, arrangement, and misfit distortion of the carbon rings that results. For instance, octagonal rings can be created with differing combinations of dimer operations so as to give differing types of neighbor rings. Adjacent STW defects can create an octagon surrounded by pentagons, hexagons and heptagons, and this is illustrated in Fig.~\ref{Octagon}(a). An octagon can also be created when a DV pinches in at left and right, as shown in Fig.~\ref{Octagon}(b). Here the resulting octagon is encircled with only pentagons and hexagons. The addition of STW defects on either side, though, results in a periphery of only hexagons, Fig.~\ref{Octagon}(c). On the other hand, two adjacent STW defects create an octagon surrounded by alternating pentagons and heptagons, and two ISTW defects create on octagon between them with a new arrangement of pentagons and heptagons, Fig.~\ref{Octagon}(d)~\cite{LuskCarr2010}. This variety of octagonal defect combinations have recently resulted in the prediction of a number of new graphene allotropes~\cite{AppelhansThesis, Appelhans2010}.

Our computational approach is outlined in Sec.~\ref{sec:methodology}, where we also provide comparative checks against previously obtained computational and experimental data. In Sec.~\ref{sec:allotropes} we consider the construction and electronic structure of metallic and semiconducting graphene allotropes. Section~\ref{sec:ribbons} focuses on the creation and analysis of thin, embedded ribbons of these allotropes within graphene to establish their structural stability and electronic character.  Finally, in Sec.~\ref{sec:conclusions} we conclude.

\section{Methodology and the Formation Energies of the Fundamental Defect Structures}
\label{sec:methodology}

Density functional theory (DFT) is employed throughout this work to estimate ground state structure and predict electronic character. A real-space numerical atomic orbital code~\cite{Delley1990} was used to initially relax the structures until the energy change was less than $2.7\times10^{-4}$ eV. A double numeric precision basis set was employed and the Perdew-Wang generalized gradient approximation used to account for electron exchange and correlation energy~\cite{Perdew1992}. This computational setting has been shown to accurately predict defect geometries and energies associated with graphene structures~\cite{LuskCarr2008, LuskCarr2009, LuskCarr2010}.

In order to more thoroughly examine the band structure of a semiconducting allotrope, a second DFT implementation was also used which employs a plane-wave basis set~\cite{VASP1, VASP2, VASP3, VASP4}. There a Projector Augmented Wave~\cite{VASPpaw1, VASPpaw2} approach with energy cutoff of 400 eV was used within a Generalized Gradient Approximation~\cite{VASP-GGA1} and 24x24x1, $\Gamma$-centered k-space grid to compute the band structure and density of states. The methodology was also extended to account for quasiparticle effects, within a $G_0W_0$ approximation,~\cite{Shishkin2007} because DFT typically underestimates the band gap of periodic structures~\cite{Payne1992}.

A STW defect was estimated to have a formation energy of 5.08 eV when embedded within a 144-atom graphene supercell; this compares well with an estimate of 4.8 eV from the literature~\cite{Li2005}. The formation energy required to create an ISTW defect, calculated using a 200-atom supercell, was found to be 6.20 eV~\cite{LuskCarr2008}. The energy required to create a DV in a 200 atom periodic supercell was 7.6 eV in agreement with previous DFT estimates of 8.7 eV~\cite{ElBarbary2003} and 7.5 eV~\cite{Kotakoski2006}. These formation energies are tabulated in Table \ref{tab:DefectEnergies}.

\begin{table}[h]
 \centering
 \caption{Defect formation energies.}
 \begin{tabular}{@{} lcc @{}} 
    \hline \\
    Defect  & eV & Number of atoms in supercell\\ \\
    \hline \hline \\
    STW        & $5.1$ & 144 \\
    ISTW       & $6.2$  & 200 \\
    DV         & $7.6$  & 200 \\
    \hline \\
    \\
 \end{tabular}
 \label{tab:DefectEnergies}
\end{table}

\section{Graphene Allotropes from Templated Defects}
\label{sec:allotropes}

\begin{figure}[hptb]
\begin{center}
\includegraphics[width=.8\textwidth]{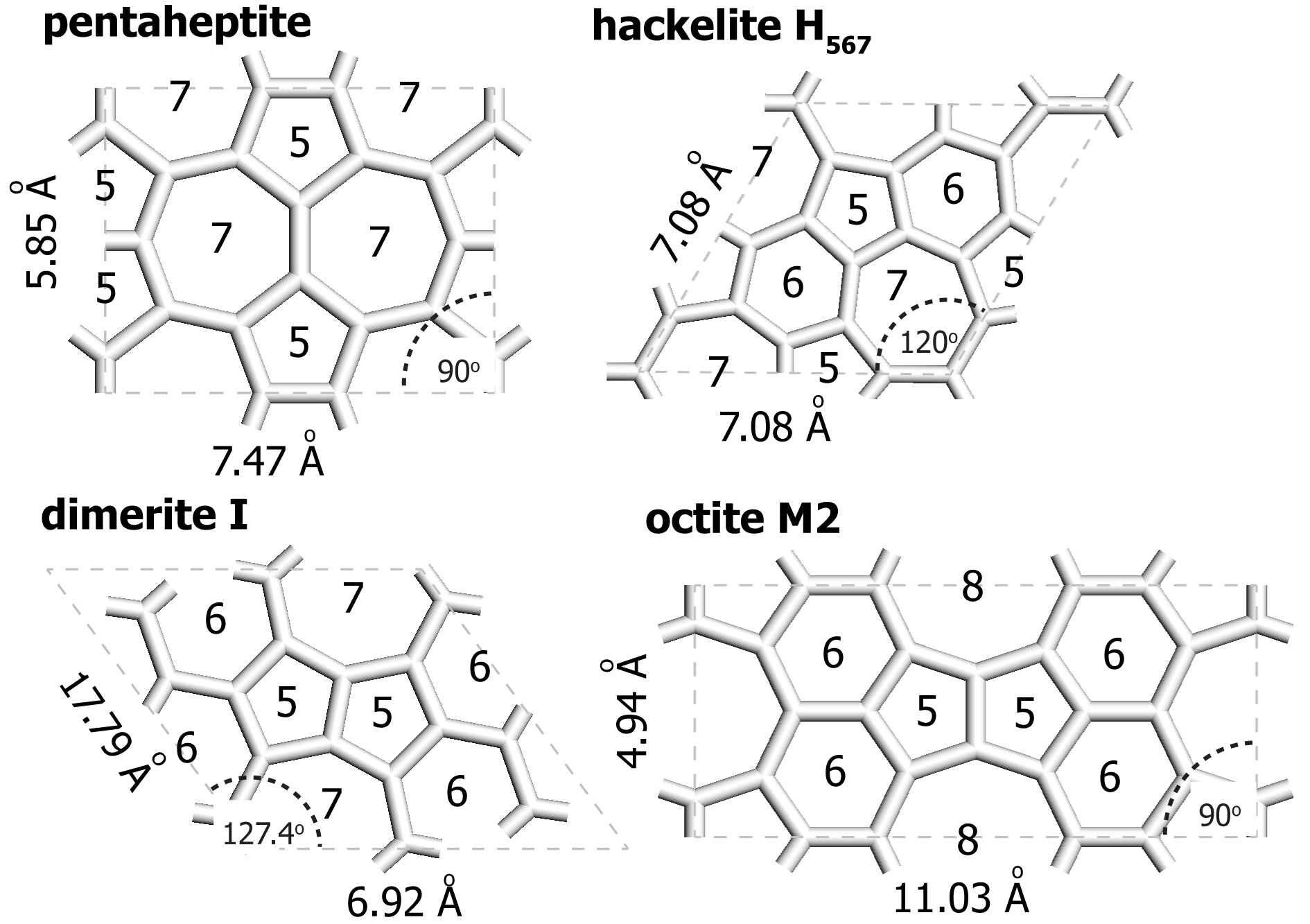}
\caption{
Representative members of graphene allotrope families: (a) pentaheptite; (b) haeckelite (c)  dimerite; (d)  octite.}
\label{AllotropeStructure}
\end{center}
\end{figure}

Even before the experimental realization of graphene in 2004~\cite{Novoselov2004}, planar allotropes had been predicted with properties distinct from graphitic sheets.  Motivated by a desire to find a pure carbon allotrope with a high Density of States (DOS) at the Fermi level, Crespi suggested that graphene allotropes with zero net curvature could form periodic, two-dimensional crystals~\cite{Crespi1996} and proposed the  pentaheptite structure shown in  Fig.~\ref{AllotropeStructure}.  A family of planar allotropes of carbon created from combinations of five-, six-, and seven-sided rings was proposed shortly thereafter~\cite{Terrones2000}. Dubbed  \emph{haeckelites}, these sheets were predicted, using tight-binding calculations, to have a metallic character. The two lowest energy structures are shown in  Fig.~\ref{AllotropeStructure}.

More recently, a tie between allotropes and defects was established by showing that  pentaheptite and all haekelites can be constructed using STW and ISTW defects~\cite{LuskCarr2009}. This led to the prediction of a new family of graphene allotropes named \emph{dimerites} because of their exclusive dependence on only the ad-dimer ISTW defect.  The lowest energy dimerite is shown in Fig.~\ref{AllotropeStructure}. While generally in higher energy than the haeckelite allotropes, dimerites are promising structures for synthesis because they can be made entirely from ad-dimers.

A subsequent focus on defect arrangements resulting in octagons led to the prediction of a new family of allotropes called \emph{octites}~\cite{AppelhansThesis, Appelhans2010}, the name reflecting the presence of eight-membered rings. The octite family includes three metallic allotropes of lower in energy than any haeckelite and an intrinsic semiconducting graphene allotrope. We use an M or SC to indicate metallic or semiconducting character.

\begin{table}[h]
 \centering
 \caption{Carbon allotropes listed according to energy. Total energy of graphene was taken as $-1037.2288$ eV/atom as computed using a $144$-atom supercell.}
 \begin{tabular}{@{} lccl @{}} 
    \hline \\
    Allotrope  & meV/atom Above Graphene&Atoms in Unit Cell& Defect Composition \\ \\
    \hline \hline \\
   graphene                    & $ 0 $ &2 & 0 \\
    octite M1       & $193$ &20 &2 STW + 2 ISTW \\
    octite M2        & $205$ &20& 2 ISTW \\
    octite M3      & $208$ &34& 2 STW + 2 ISTW \\
    haeckelite $H_{567}$             & $233$ &16 & 1 STW + 1 ISTW\\
    dimerite I                  & $284$ &16& 1 ISTW \\
    octite SC    & $318$ &28 & 2 STW + 1 DV \\
    dimerite II                 & $330$ &40& 2 ISTW \\
    octite M4           & $357$ &8& 1 STW \\
    haeckelite $O_{567}$             & $364$ &12& 1 STW \\
    dimerite III                & $366$ &30& 3 ISTW \\
    haeckelite $R_{57}$              & $372$ &16& 2 STW \\
    buckeyball C$_{60}$         & $372$ &60& - \\
    \hline \\
    \\
 \end{tabular}
 \label{tab:StructureEnergies}
\end{table}

\begin{figure*}[hptb]
\begin{center}
\includegraphics[width=0.9\textwidth]{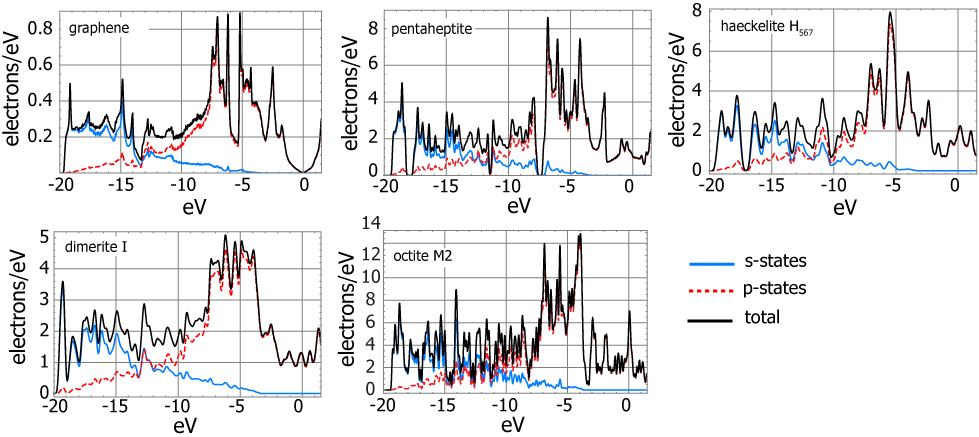}
\caption{
Density of states of graphene, pentaheptite, haeckelite H$_{567}$, dimerite I and octite M2. All four allotropes are metallic. A (12, 6, 12) Monkhorst-Pack grid was employed along with a linear interpolation grid size of 0.05 eV.}
\label{AllotropeDOS}
\end{center}
\end{figure*}

The pentaheptite, haeckelites and dimerites are predicted to be conductors as shown in the Density of States (DOS) summaries of Fig.~\ref{AllotropeDOS}. All of these can be constructed by patterning of STW and ISTW defects. Similarly, metallic octites are composed of STW and ISTW defects. This apparent correlation motivated us to search for a new type of defect building block that would lead to a semiconducting carbon sheet. The result was the addition of the DV defect as a third basic building block~\cite{Appelhans2010}.

\begin{figure}[ptb]
\begin{center}
\includegraphics[width=0.9\textwidth]{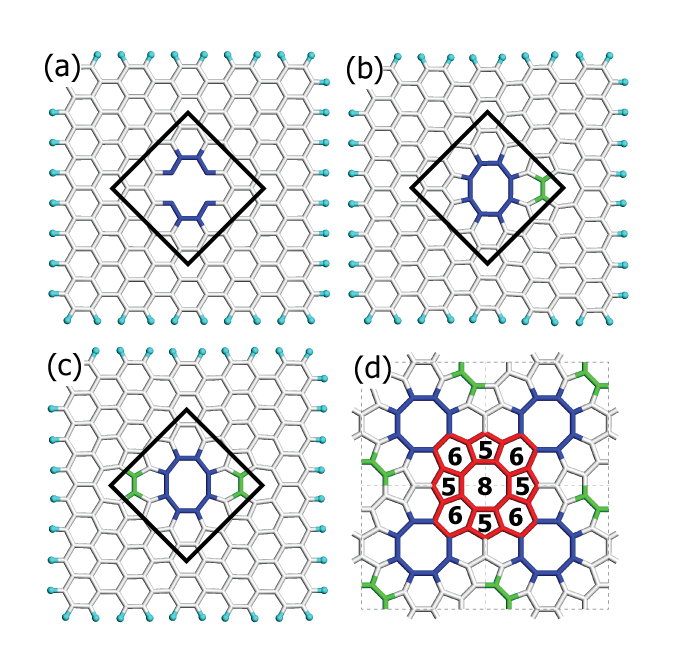}
\caption{A patch of semiconducting octite constructed on passivated graphene: (a) graphene with a di-vacancy; (b) an STW defect is introduced by rotating the gray (green) bond; (c)  a second STW defect is introduced; (d) patterning of the patch results in a second octagonal defect surrounded by alternating hexagons and pentagons, shown in gray (red).}
\label{Synthesis}
\end{center}
\end{figure}

Of particular interest is a new semiconducting allotrope, octite SC, which can be constructed from a combination of STW and DV defects. It amounts to a patterning of the defect assemblies shown in Fig.~\ref{Octagon} (a) and (b). To see this, consider the 170-atom carbon sheet of Fig.~\ref{Synthesis}(a). The edges are passivated with hydrogen to maintain sp-2 bonding for all carbon atoms and a single DV defect is introduced.  Two bond rotation defects are subsequently imposed (b and c). The result is then replicated so as to create an array of octagonal defects. Octagons surrounded by hexagons result from DV and STW defects while smaller octagons surrounded by alternating pentagons and hexagons derive from STW defects (d).  The planar arrangement of atoms with three nearest neighbors suggests sp-2 bonding, although variations in bond length (1.40 $\AA$ to 1.54 $\AA$) indicate that the three-fold symmetry of graphene bonding is no longer present. For the sake of comparison, sp-2 bond lengths in graphene are 1.42 $\AA$ while the sp-3 bond length of diamond are 1.54 $\AA$.

\begin{figure}[hptb]
\begin{center}
\includegraphics[width=0.9\textwidth]{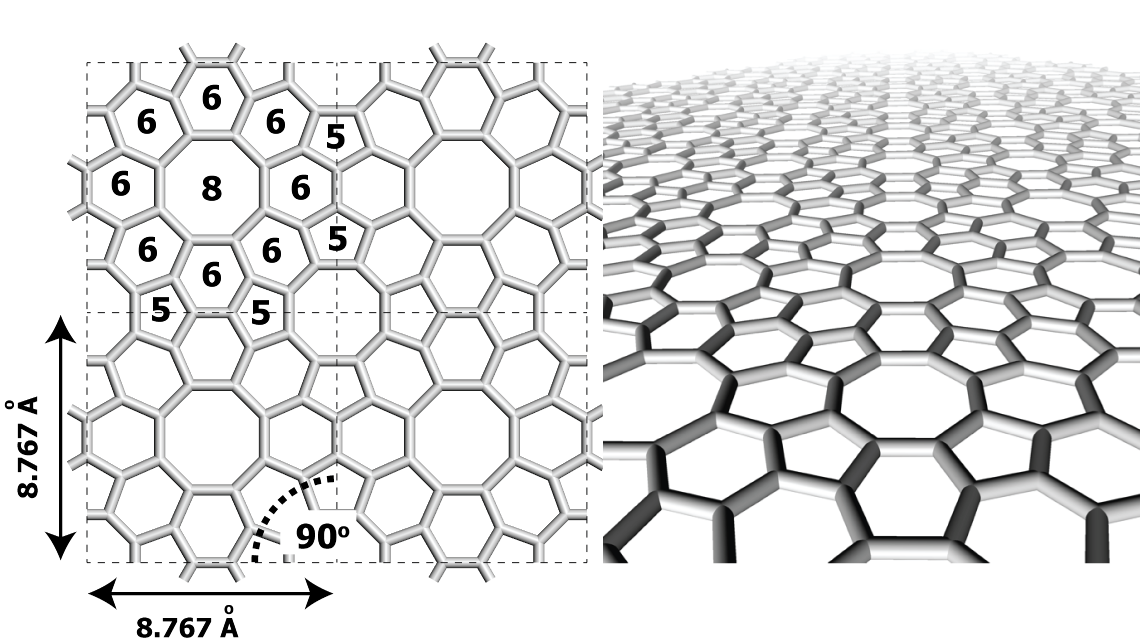}
\caption{
A computationally created semiconducting allotrope of graphene resulting from patterned defects. The planar structure is only 313 meV/atom above graphene, comparable to 233 meV/atom for haeckelite $H_{567}$, a metallic graphene allotrope previously theorized~\cite{Terrones2000}: (left) 4 unit cells showing arrangements of rings around a central octagon; (right) perspective view making clear that this allotrope is planar.}
\label{SCgeometry}
\end{center}
\end{figure}

The stand-alone form of this semiconductor is shown in Fig.~\ref{SCgeometry}. It has a primitive cell of 28 atoms arranged in pentagons, hexagons, and octagons. The planar density is 0.364~atoms per $\AA^2$, comparable to 0.380~atoms per $\AA^2$ for graphene. The central geometrical feature is an octagon completely surrounded by hexagons. The cell geometry is square with p4/MMM symmetry. A Hessian linear vibrational analysis showed the structure to be a local energy minimum. Room temperature quantum molecular dynamics simulations were also used to confirm its stability.

\begin{figure}[ptb]
\begin{center}
\includegraphics[width=.9\textwidth]{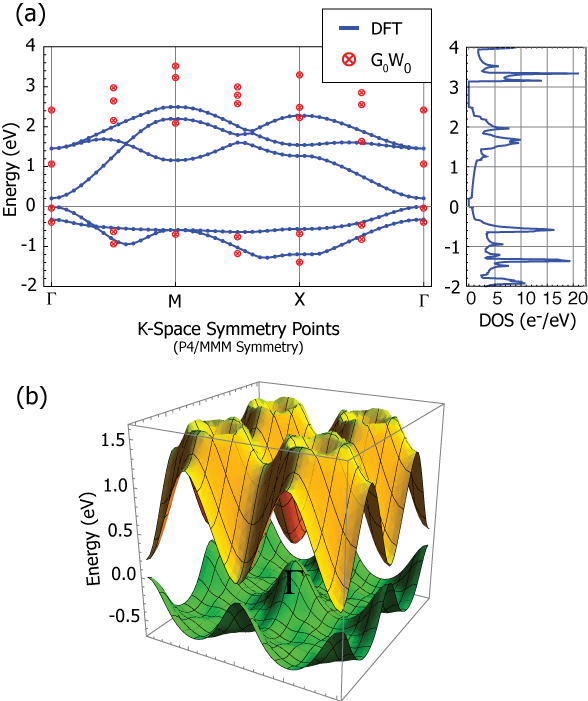}
\caption{Electronic structure of the octite semiconductor: (a) band structure (BS) and density of states (DOS) for a semiconducting graphene allotrope. Both DFT and $G_0W_0$ predictions are shown. The DOS plot is from the DFT calculation; (b) perspective view of the band structure over the entire two-dimensional k-space.}
\label{BDandDOS}
\end{center}
\end{figure}

The ground state structure was verified using a second DFT implementation that employs a plane-wave basis set~\cite{VASP1, VASP2, VASP3, VASP4}. The wave function energy cutoff was set at 400 eV, and a conjugate gradient method employing a 4x4x2 Monkhorst-Pack grid resulted in the same ground state structure but with a lattice constant of 8.674 $\AA$. The plane-wave DFT code was used to predict both the band structure (BS) and Density of States (DOS). A 24x24x1, $\Gamma$-centered k-space grid was used to calculate the electronic properties. The BS and DOS diagrams are shown in  Fig.~\ref{BDandDOS}. The material possesses a 0.2 eV direct band gap at the $\Gamma$ point. As in graphene, all carbon atoms bond with three nearest neighbors, but the symmetry of the in-plane sp-2 character has been disrupted by the varying bond angles and bond lengths.

DFT typically underestimates the band gap of periodic structures~\cite{Payne1992}, so the 0.2 eV gap should be viewed as a lower bound. Many-body perturbation theory, within the $G_0W_0$ approximation~\cite{Shishkin2007}, was therefore used to provide a more accurate estimate of the electronic structure. As expected, the inclusion of quasiparticle effects widens the band gap, providing an estimate for the direct, $\Gamma$ point gap of 1.1 eV. The $G_0W_0$ band structure is shown with its DFT counterpart in  Fig.~\ref{BDandDOS}. Three iterations towards self-consistency ($GW_0$ approximation) opened the band gap slightly to 1.2 eV.

Metallic allotropes can also be composed of octagonal rings, and three have been computationally predicted to have the lower ground state energies than any previous allotrope~\cite{AppelhansThesis}. One of them, octite M1, has an energy only 194 meV/atom above graphene--lower than the 233 meV/atom calculated for the lowest energy haeckelite or the 284 meV/atom for the lowest energy dimerite. This structure is important because it is expected to be the most stable of any predicted graphene derivative which therefore makes it a likely candidate for experimental realization. In all of the metallic octites, octagonal rings are built exclusively from  ISTW and STW defects. This is in contrast to the semiconducting octite wherein DVs are employed to build octagons with a different local environment.

\begin{figure}[ptb]
\begin{center}
\includegraphics[width=.7\textwidth]{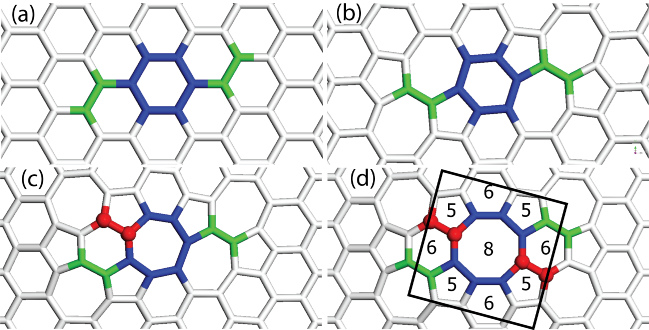}
\caption{Defect engineering a patch of octite M1 within graphene: (a) graphene with highlights on atoms to be manipulated; (b) 2 STW bond rotations; (c) first ISTW ad-dimer; (d) second ISTW ad-dimer.}
\label{DVMConstruction}
\end{center}
\end{figure}

From the perspective of defect engineering, the octite M1 can be constructed from a template composed of two STW defects and two ISTW defects as shown in Fig.~\ref{DVMConstruction}. This synthesis route was carried out computationally on a rectangular supercell of graphene measuring 21.34 by 12.32 $\AA$. The intermediate steps were geometrically optimized DFT ground states after each defect was introduced, but the lattice was not allowed to dilate. The initial zero energy state is graphene with two dimers at infinity. Two STW bond rotations (b) raise the energy to 9.1 eV but dimer addition through ISTW defects lowers the energy to 1.6 eV (c) and finally to -6.8 eV.
%

\begin{figure}[ptb]
\begin{center}
\includegraphics[width=0.9\textwidth]{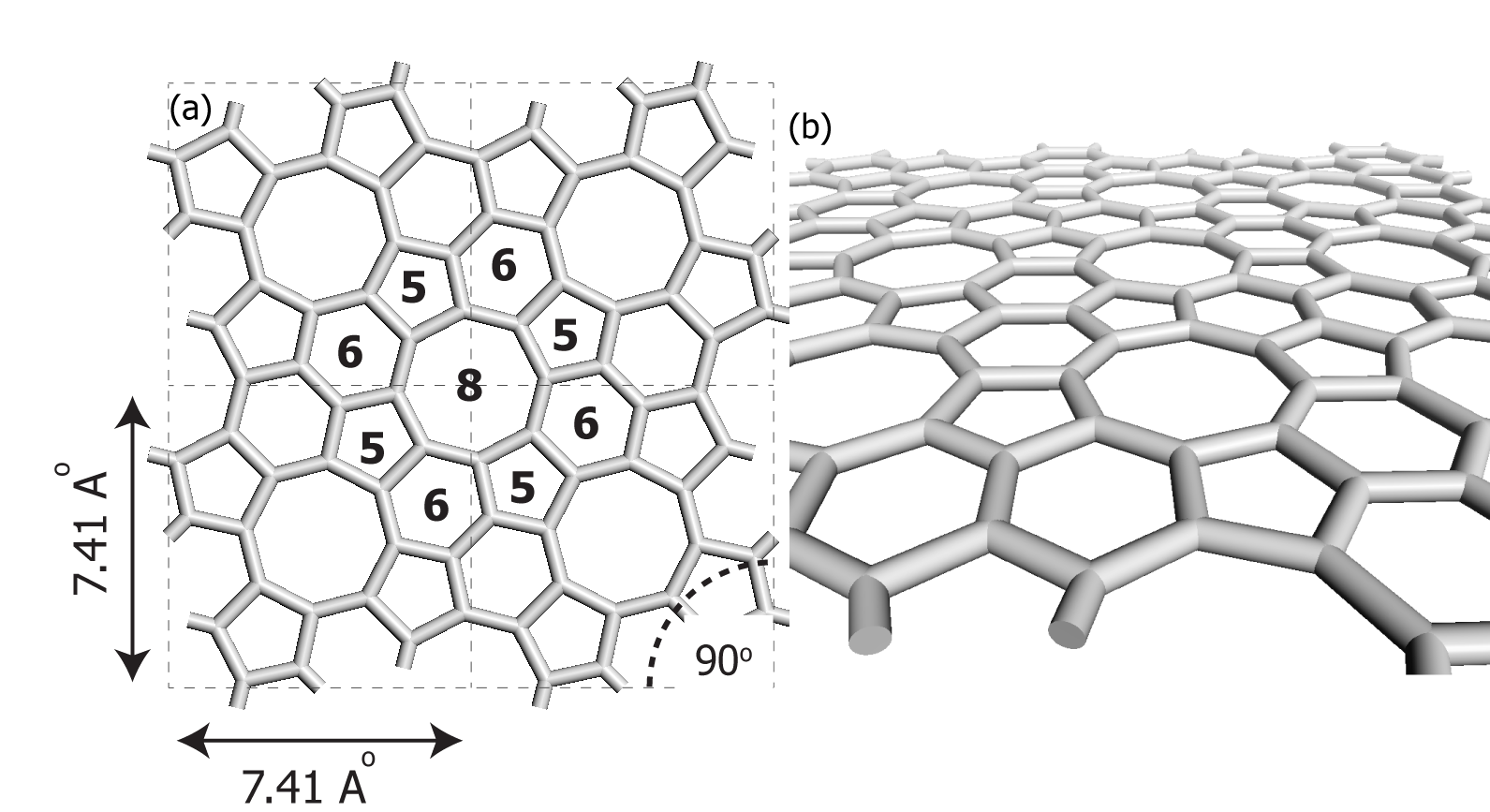}
\caption{2x2 supercell of the lowest energy octite M1: (a) top view; (b) perspective view.}
\label{DVMGeometry}
\end{center}
\end{figure}

The resulting octite M1, shown in  Fig.~\ref{DVMGeometry}, is completely flat and contains 20 atoms in the primitive unit cell. The energy minimum was verified via a Hessian vibrational matrix. Quantum molecular dynamics simulations were also utilized, at 300 K and 500 K, to ensure that the structure is indeed stable.

\begin{figure}[ptb]
\begin{center}
\includegraphics[width=0.9\textwidth]{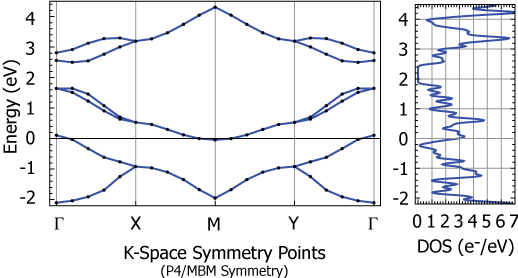}
\caption{Band structure and total density of states (DOS) for the lowest energy graphene allotrope, octite M1. The DOS is dominated by the p-state contribution.}
\label{DVMElectrical}
\end{center}
\end{figure}

Density of States (DOS) calculations were performed using a Monkhorst-Pack~\cite{Monkhorst1976} grid size of 12x12x4 which was then interpolated onto a grid of 200x200x200 points. A slight broadening of 0.05 eV was applied for a more physical representation of the DOS in a large system.  Fig.~\ref{DVMElectrical} shows that octite M1 is a good conductor with approximately 2 electrons/eV at the Fermi level.

\section{Embedded Ribbons of Graphene Allotropes}
\label{sec:ribbons}

The link between defects and allotropes provides the design guidance needed to create both localized and extended regions of semiconducting and/or conducting material within graphene. We take up this issue by investigating embedded allotropic ribbons, i.e., a thin strip of one allotrope embedded within another. These are distinguished from nanoribbons which are thin strips of material with free boundaries. Nanoribbons have been promoted as a means of endowing graphene with a tunable band gap, and the approach requires that a thin strip of graphene be precisely cut to a prescribed chirality and width. Lateral confinement of the charge carriers creates a band gap~\cite{Han2007}. Computational studies have determined that nanoribbons can be conductive or insulating depending on the shape of their edges~\cite{Brey2006}. Embedded allotropic ribbons offer an alternative strategy wherein both conducting and semiconducting pathways can be created on a single, planar sheet of carbon. We examine both types of ribbons embedded within a graphene sheet.

\begin{figure*}[ptb]
\begin{center}
\includegraphics[width=0.9\textwidth]{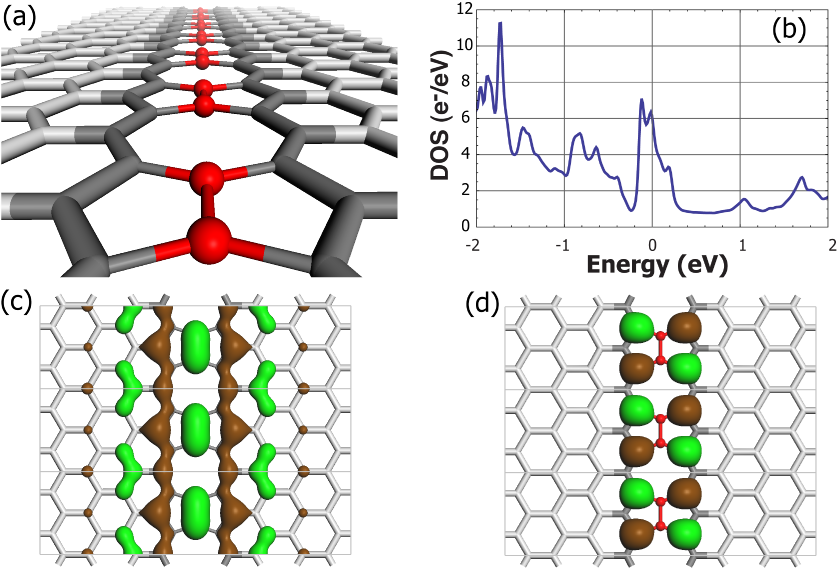}
\caption{Minimal ribbon of octite M2 within graphene: (a) Perspective view of the extended defect; (b) DOS exhibiting good conduction with 20 meV interpolative instrument broadening~\cite{Ackland1998}; (c) valence band edge state; (d) conduction band edge state. Isosurfaces are for 0.03 electrons/\AA$^3$. Green and brown coloring distinguishes regions for which the orbital wave function has positive and negative phase. The orbitals of panels (c) and (d) are constructed using a $\Gamma$-point ($k=0$) spatial frequency.}
\label{DimeriteStrandResults}
\end{center}
\end{figure*}

A minimal ribbon of a conducting octite (Figs.~\ref{AllotropeStructure} and \ref{AllotropeDOS}) within graphene is shown in Fig.~\ref{DimeriteStrandResults}(a).  Within this setting, the extended defect amounts to an alternating sequence of octagons and paired pentagons. The relaxed, planar ground state structure is a conductor (Fig.~\ref{DimeriteStrandResults}(b)) with band edge states consistent with that previously determined for isolated ISTW defects~\cite{LuskCarr2010}. Specifically, the tops of the ad-dimers repel additional electrons while the shoulders of the extended defect actually attract additional electrons. This is made clear in Fig.~\ref{DimeriteStrandResults}(c, d) which show the highest occupied and lowest unoccupied molecular orbitals (HOMO/LUMO). These orbitals are constructed using a $\Gamma$-point ($k=0$) spatial frequency, i.e., by setting the Bloch function wave number to zero and relying on linear combinations of atomic orbitals to describe the electronic structure. These band edge states are both localized along the ribbon, and the ribbon itself endows the composite with a high DOS at the Fermi level, similar to vacancy localization effects found in bilayer graphene\cite{Castro2010}. The composite structure is a conductor as shown in Fig.~\ref{DimeriteStrandResults}(b).

\begin{figure}[hptb]\begin{center}
\includegraphics[width=0.9\textwidth]{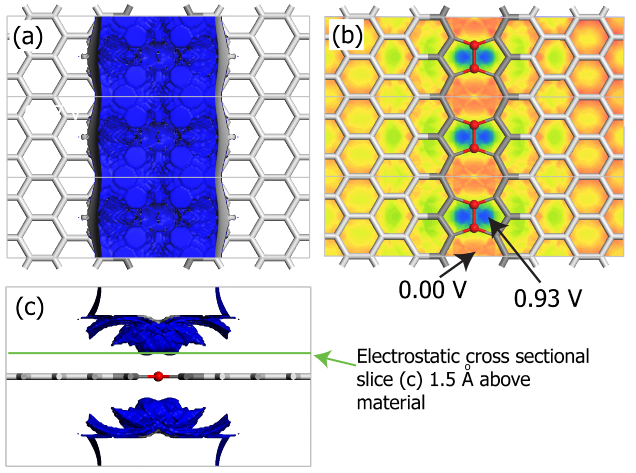}
\caption{
Electrostatic field of the minimal ribbon of octite M2 shown in Fig.~\ref{DimeriteStrandResults}(a): (a) isosurface of 0.27 V; (b) cross-section at 1.7 $\AA$  above surface ranging from 0.0 V (red) to 0.73 V (blue); (c) side view of isosurface showing spatial orientation of the cross-sectional plot.}
\label{fig:ISTWribbon_electrostatic}
\end{center}
\end{figure}

The conducting nature of the composite is presumably due to electron transport along the extended defect. This is supported by plots of the electrostatic potential provided in Fig.~\ref{fig:ISTWribbon_electrostatic}. The isosurface shown in panel (a) indicates that the defect has a positive potential and thus attracts additional charge, while panel (b) shows potential contours in a plane that intersects the isosurface.

\begin{figure}[hptb]\begin{center}
\includegraphics[width=0.48\textwidth]{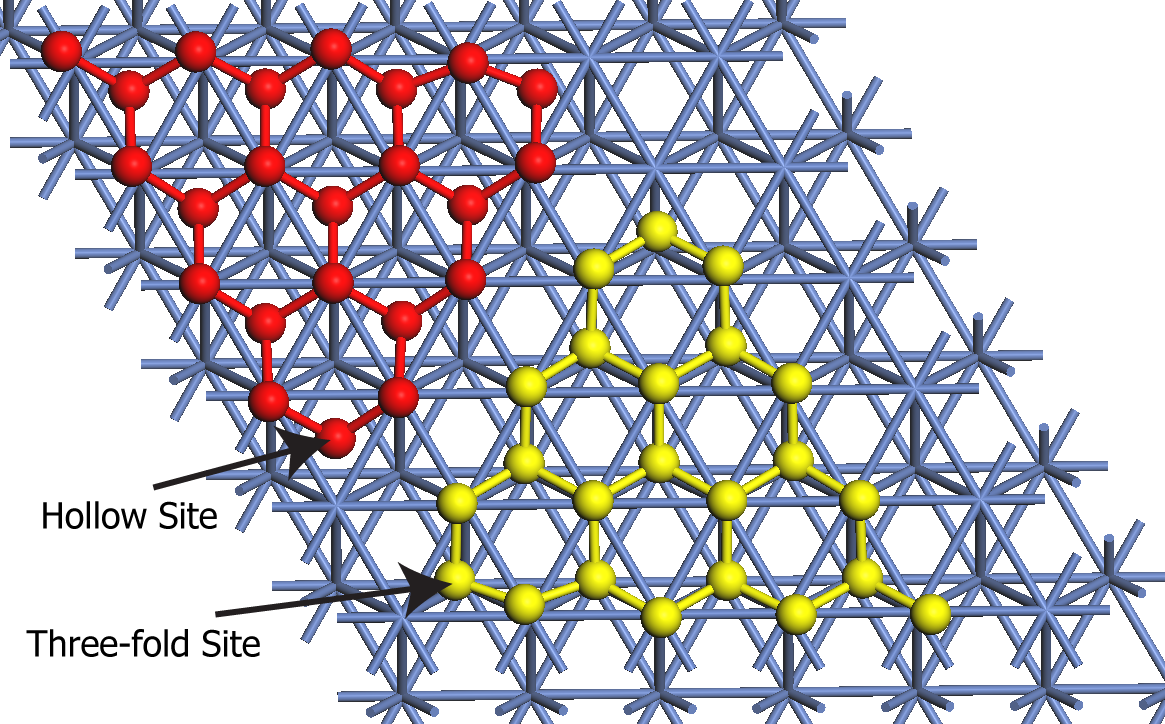}
\caption{
Ni(111) surface (nickel lattice shown in blue) is used to grow flakes of graphene centered on either three-fold (yellow carbon atoms) or hollow (red carbon atoms) sites. The extended defect where dissimilar flakes meet is the minimal ribbon of metallic octite shown in Fig.~\ref{DimeriteStrandResults}(a).}
\label{fig:ISTWribbon_Batzill_schematic}
\end{center}
\end{figure}

\begin{figure}[ptb]
\begin{center}
\includegraphics[width=0.45\textwidth]{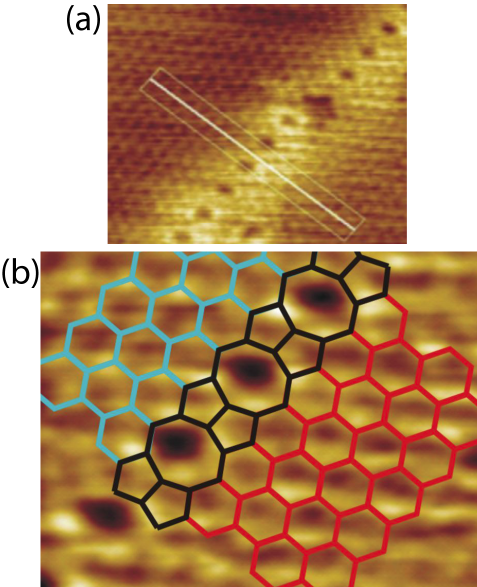}
\caption{Experimentally realized ribbon of octite M2 courtesy of Ref.~\cite{Batzill2010}. The structure is identical to the minimal ribbon of octite shown in Fig.~\ref{DimeriteStrandResults}: (a) Perspective view showing extended defect; (b) close-up view of region outlined in white in panel (a) with overlay schematic of lattice structure.}
\label{BatzillStrand}
\end{center}
\end{figure}

Significantly, this metallic defect ribbon has recently been experimentally synthesized~\cite{Batzill2010, CarrLusk2010, AppelhansThesis}. Graphene was grown epitaxially on the (111) surface of nickel, where graphene may be aligned in one of two  nearly energetically equivalent  ways: one-half of the carbon atoms may lie over either three-fold or hollow sites as shown in  Fig.~\ref{fig:ISTWribbon_Batzill_schematic}. The mismatch in carbon atoms at the interface of dissimilar graphene flakes results in the extended defect structure of Fig.~\ref{DimeriteStrandResults}. The nickel can be subsequently removed resulting in a stand-alone sheet of graphene with a minimal ribbon of dimerite. Although not measured experimentally, DFT analysis performed by Batzill \textit{et al.}~\cite{Batzill2010} in the localized density approximation suggested that these ribbons are metallic, consistent with our own DFT studies using a generalized gradient approximation. The experimental result is provided in Fig.~\ref{BatzillStrand}~\cite{Batzill2010}. This suggests that the realization of more complex defect designs is only a matter of time.

%

\begin{figure*}[ptb]
\begin{center}
\includegraphics[width=0.9\textwidth]{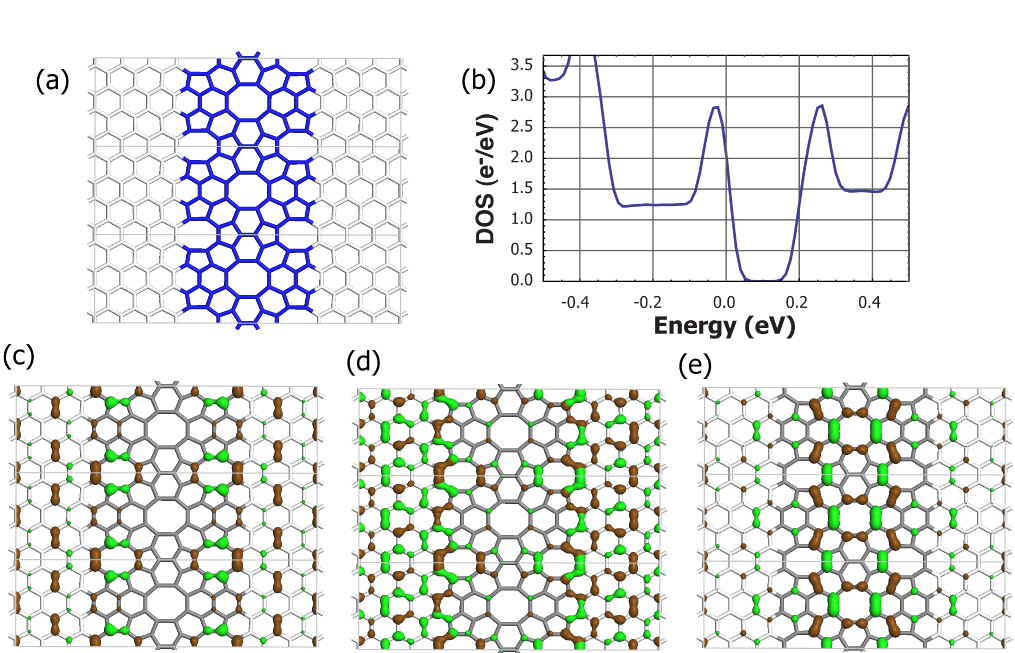}
\caption{A ribbon of octite SC built within graphene: (a) three periodic cells showing interfacial structure; (b) DOS with interpolative instrument broadening of 20 meV~\cite{Ackland1998}; (c) valence band edge state; (d) conduction band edge state; (e) LUMO+1 state. Isosurfaces are for 0.03 electrons/\AA$^3$. Green and brown coloring distinguishes regions for which the orbital wave function has positive and negative phase. The orbitals of panels (c) and (d) are constructed using a $\Gamma$-point ($k=0$) spatial frequency.}
\label{SC_strand_4_results}
\end{center}
\end{figure*}

Embedded ribbons of semiconducting octite (Fig.~\ref{SCgeometry}) were subsequently investigated. The dimensions of the periodic cell were fixed to simulate a constraining lattice of graphene, and DFT was used to obtain the planar, ground state structure shown in  Fig.~\ref{SC_strand_4_results}(a). The DOS plot of Fig.~\ref{SC_strand_4_results}(b) indicates that the ribbon endows the graphene with a 0.17 eV band gap, only slightly smaller than the 0.2 eV band gap of the octite SC itself.  Unlike the metallic ribbon,  the HOMO and LUMO states are not confined to the ribbon, although they do have contributions from orbitals along the interfaces between the two allotropes (Fig.~\ref{SC_strand_4_results}(c, d)). Localized states appear along the edge of zigzag nanoribbons~\cite{Kyoko1996} and interestingly, the LUMO+1 state is highly localized within the ribbon as shown in Fig.~\ref{SC_strand_4_results}(e).

\begin{figure}[hptb]\begin{center}
\includegraphics[width=0.9\textwidth]{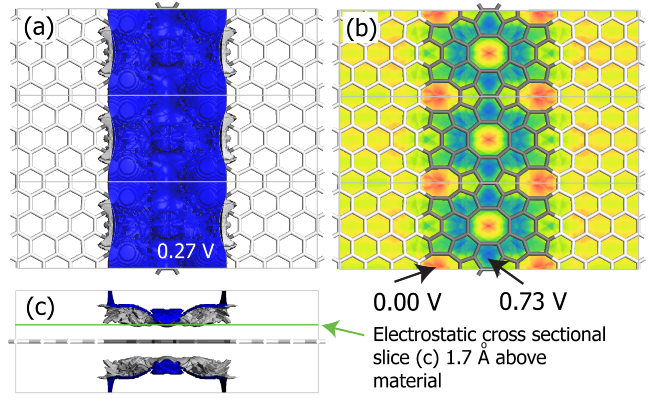}
\caption{
Electrostatic field of octite SC forming a diamond network: (a) isosurface of 0.27 V; (b) cross-section at 1.7 $\AA$ above surface ranging from 0.0 V (red) to 0.73 V (blue).}
\label{fig:SC_strand_4_electrostatic}
\end{center}
\end{figure}

The electronic structure of the semiconducting ribbon is further analyzed by plotting the electrostatic field. As shown in Fig.~\ref{fig:SC_strand_4_electrostatic}, potential isosurfaces tend to be more localized on the ribbon as the distance above the sheet increases. These potential surfaces are positive, indicating that additional charge will be attracted to the semiconducting ribbons.
%

\begin{figure*}[hptb]\begin{center}
\includegraphics[width=0.9\textwidth]{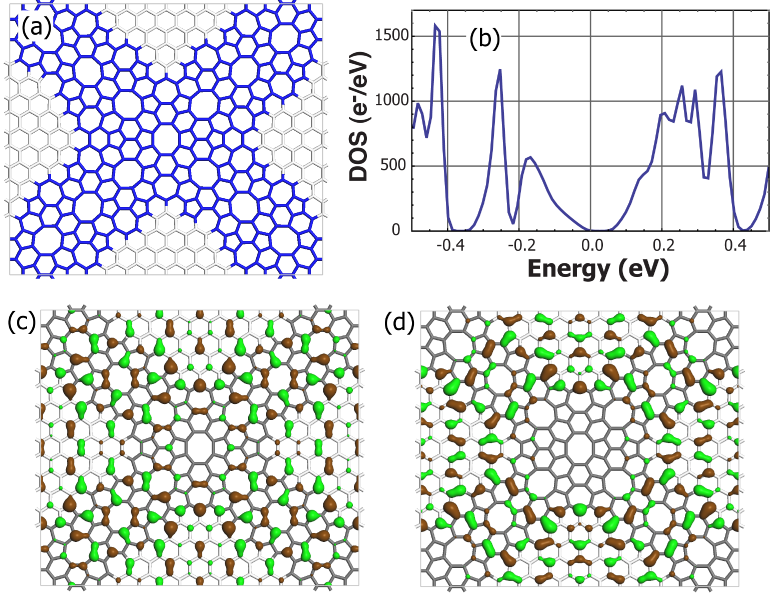}
\caption{
(a) Embedded ribbons of semiconducting octite form a network partitioning graphene into diamond-shaped domains.  (b) DOS with 10 meV interpolative instrument broadening~\cite{Ackland1998} shows that the ribbons disrupt electron conduction.  (c) Valence band edge state and (d) conduction band edge state. Isosurfaces are for 0.01 electrons/\AA$^3$. Green and brown coloring distinguishes regions for which the orbital wave function has positive and negative phase. The orbitals of panels (c) and (d) are constructed using a $\Gamma$-point ($k=0$) spatial frequency.}
\label{fig:SC_X_results}
\end{center}
\end{figure*}

Embedded ribbons may be constructed with a variety of chiralities, and this can be exploited to create junctions and ribbon networks. Two examples of such semiconductor networks are considered. The first is shown in  Fig.~\ref{fig:SC_X_results}(a), where the enforcement of periodic boundary conditions gives graphene islands in the shape of diamonds. For this orientation of ribbons, the DOS plot indicates that the sheet still maintains a small band gap (~0.05 eV). The HOMO state is localized within the ribbons with the exception of the ribbon junction, while the LUMO state shows that the first excited state will tend to avoid the junction completely.

\begin{figure}[hptb]\begin{center}
\includegraphics[width=0.48\textwidth]{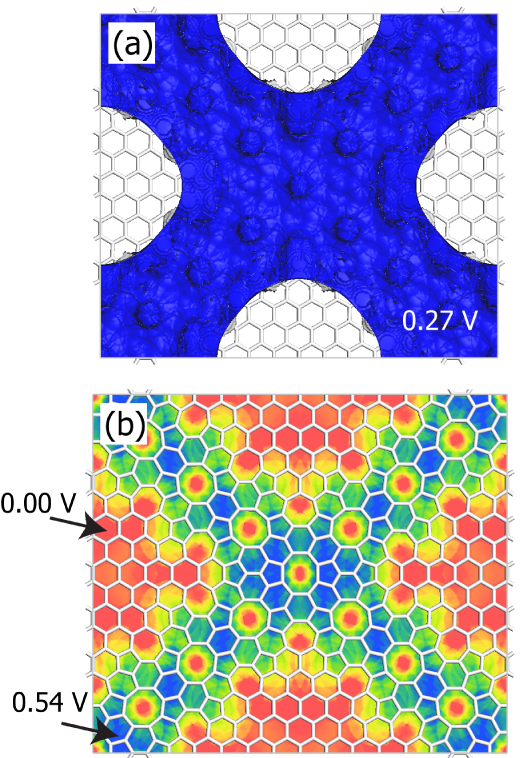}
\caption{
Electrostatic field of DVSC form a diamond network: (a) isosurface of 0.27 V; (b) cross-section at 1.5 $\AA$ above surface ranging from 0.0 V (red) to 0.54 V (blue).}
\label{fig:SC_X_electrostatic}
\end{center}
\end{figure}

The electrostatic potential of the diamond network is shown in  Fig.~\ref{fig:SC_X_electrostatic}. Consistent with the results for isolated allotrope interfaces, potential surfaces lying furthest above the sheet are positive and localized on the ribbon network. Higher energy excited electrons should therefore be attracted to these conduction channels.

%

\begin{figure*}[hptb]\begin{center}
\includegraphics[width=0.9\textwidth]{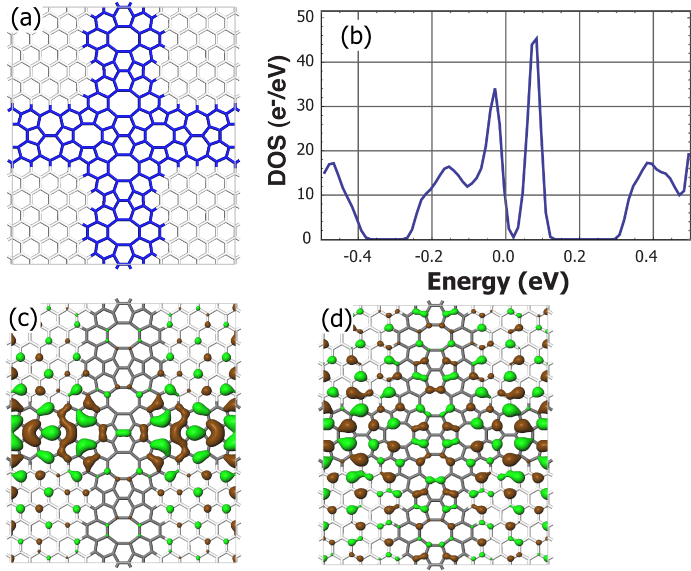}
\caption{
(a) Embedded ribbons of semiconducting octite form a network partitioning graphene into rectangular domains.  (b) DOS with 10 meV interpolative instrument broadening~\cite{Ackland1998} shows that the ribbons disrupt electron conduction. (c) Valence band edge state and (d) conduction band edge state. Isosurfaces are for 0.015 electrons/\AA$^3$. Green and brown coloring distinguishes regions for which the orbital wave function has positive and negative phase. The orbitals of panels (c) and (d) are constructed using a $\Gamma$-point ($k=0$) spatial frequency.}
\label{fig:SC_T_results}
\end{center}
\end{figure*}

Perhaps not surprisingly, the orientation of intersecting ribbons impacts the electronic structure of the sheet. This is made clear by considering a second network of semiconducting ribbons which now partitions the graphene into rectangular islands. Figure~\ref{fig:SC_T_results}(a) shows the planar, ground state geometry. The DOS of this system (panel b) indicates that the band gap is essentially lost. The HOMO state is now localized along the horizontal semiconducting ribbon (panel c), but the two states of immediately lower energy do not show localization along either ribbon orientation. This suggests that the two orientations have fundamentally different electronic properties which likely extends to their respective electron mobilities.

\begin{figure}[hptb]\begin{center}
\includegraphics[width=0.48\textwidth]{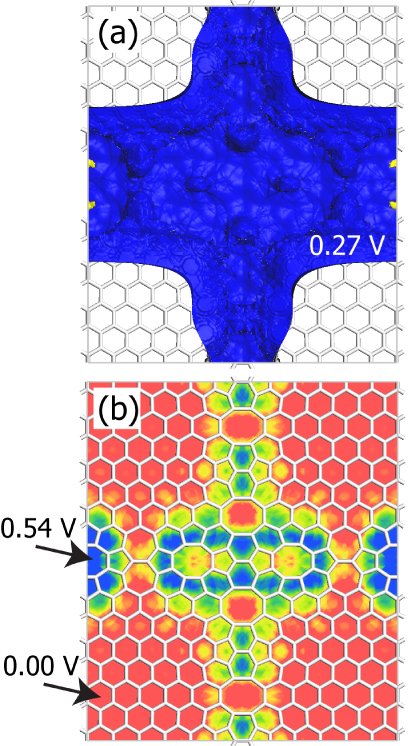}
\caption{
Electrostatic field of octite SC forming a rectangular network: (a) isosurface of 0.27 V; (b) cross-section at 1.5 $\AA$ above surface ranging from 0.0 V (red) to 0.54 V (blue).}
\label{fig:SC_T_electrostatic}
\end{center}
\end{figure}

The electrostatic potential of the rectangular network, shown in  Fig.~\ref{fig:SC_T_electrostatic}, is very similar to that of the diamond networked sheet. Once again, potential surfaces lying furthest above the sheet are positive and localized on the ribbon network. Higher energy excited electrons should therefore be attracted to these conduction channels.  The electrostatic contours of  Fig.~\ref{fig:SC_T_electrostatic}(c), though, show that the horizontal  ribbon attracts electrons more strongly at a fixed height above the sheet.

%

\section{Conclusions}
\label{sec:conclusions}
Defect engineering offers a means of changing the electronic character of graphene without the need to introduce new atomic elements. An algebraic framework has been identified for the modification of lattices and is based on four basic actions which can be performed with a carbon dimer:  bond rotation (STW defect);  dimer addition (ISTW defect); dimer removal (divacancy defect); and no action (standard graphene dimer). These dimer actions can be combined to build a variety of localized defect structures;  the current work has focused on the templating of elements to create extended defects. Two-dimensional templates can be used to create graphene allotropes. While most of these are good conductors, we have identified a defect structure which opens up a band gap of 1.2 eV. These allotropes can be created as conducting and semiconducting ribbons within graphene, extended one-dimensional defect structures. Their electronic character is largely preserved for even the thinnest ribbons,  which offers the prospect of designing electronic components crafted entirely out of carbon.

\section{Acknowledgements}
We are pleased to acknowledge the use of computing resources provided through the Golden Energy Computing Organization (NSF Grant No. CNS-0722415) and the Renewable Energy MRSEC program (NSF Grant No. DMR-0820518) at the Colorado School of Mines.  LDC thanks the Institute for Advanced Study at Tsinghua University where part of this work was completed.



\end{document}